# An Insider Threat Mitigation Framework Using Attribute Based Access Control

Olusesi Balogun*
Georgia State University
Atlanta, GA 30303 USA
obalogun6@student.gsu.edu

Daniel Takabi
Georgia State University
Atlanta, GA 30303 USA
takabi@gsu.edu

## ABSTRACT

Insider Threat is a significant and potentially dangerous security issue in corporate settings. It is difficult to mitigate because, unlike external threats, insiders have knowledge of an organization's access policies, access hierarchy, access protocols, and access scheduling. In addition, the complexity, time, and skill required to locate the threat source, model, and timestamp make it more difficult for organizations to combat. Several approaches to reducing insider threat have been proposed in the literature. However, the integration of access control and moving target defense (MTD) for deceiving insiders has not been adequately discussed. In this paper, we combine MTD, deception, and attribute-based access control to make it more difficult and expensive for an insider to gain unauthorized access. We introduce the concept of correlated attributes into ABAC and extend the ABAC model with MTD by generating mutated policy using the correlated attributes for insider threat mitigation. The evaluation results show that the proposed framework can effectively identify correlated attributes and produce adequate mutated policy without affecting the usability of the access control systems.

## 1 INTRODUCTION

Insider Threat is an existent and significant security challenge in corporate organizations. Recently, insider threat has increasingly grown to be a leading threat as more organizations adopt digital tools to manage their data, either locally or in the cloud, resulting in expensive security breaches. [13] According to Cyber and Infrastructure Security Agency (CISA), "insider threat occurs when an insider uses authorized access, wittingly or unwittingly, to do harm to the Department's mission, resources, personnel, facilities, information, equipment, networks, or systems. [6]" Organization's employees misuse their assigned privileges to fulfill specific secondary purposes. These purposes include gaining access to either view, edit, delete unauthorized resources, or reveal the resources to the organization's competitors. Unlike external threat, which organizations often guard against mainly at the network level before entering their perimeter, insider threat lingers for an extended period and mostly goes unnoticed before its effect surfaces. In addition, it is difficult to manage because, unlike external threats, the insiders have some information about an organization's access policies, access hierarchy, access protocols, and access scheduling. In addition, the complexity, time, and expertise required to identify the threat's source, technique, and timeline make it more challenging for organizations to combat. Therefore, insider threat remains a top security priority for private and public organizations. In this light, some works have numerically quantified the severity of the insider threat to organizations. For example, in 2015, a survey of federal cybersecurity containing 200 Information Technology managers conducted by the SolarWinds research team revealed that one-third of the participants had concerns about insider threats. [24]. Likewise, in 2020, the Securonix Threat Research Team examined over 300 security incidents across organizations from 8 different sectors. According to their findings, privilege abuse, a key type of insider threat, accounts for 19% of all incidents and is the second most common type. [23] More recently, Cybersecurity Insiders Research Team conducted an insider threat study of organizations. Their result revealed that 98% of organizations are vulnerable to insider attacks, while 49% of such organizations only detect insider threats after the data have been copied out of the organization [12].

In the literature, the proposed approaches to mitigate insider threat can generally be classified into four lines of research which are: psychological-driven, behavioral-driven, content-driven, and deception-driven. The psychological-driven approach tracks insider threats using human signals such as brain signals [10], and eye movements [9]. The behavioral-driven approach performs data analytics of human behavior using methods such as analysis of access logs [15] or convolution graph [26]. The content-driven approach builds natural language and machine learning models on texts to detect insider threats. [5, 20]. Lastly, deception-driven approach uses honey items such as honey attributes [1], honeypots [3], honey files [27], honey permissions [17], honey words[16], honey documents[4], honey tokens [2, 25], and honey encryption[18] to deceive insiders. Further, several studies have integrated insider threat mitigation approaches to access control systems, such as Attribute-based Access control (ABAC), to monitor insiders to achieve better mitigation. ABAC is an access control system that authorize or denies user's access request to an object based on the user's attributes, object's attributes, environment's attributes, and policy rules. ABAC has emerged as a promising access control system and is thus commonly used within and across organizations because: (1) it can be extended seamlessly into other frameworks for better security enhancement; (2) its components can flexibly be adjusted or increased; (3) it is granular and can be managed easily; and (4) it can effectively handle concurrency of users. However, a possible security breach to ABAC can occur if an insider has information about the policy rules and changes them to gain unauthorized access. ABAC's entities can be changed randomly to mitigate this incident based on heuristics such as Moving Target Defense (MTD). MTD is a potential security concept in the literature that proactively changes security systems' configurations to deter potential attacks. This work presents ongoing research integrating ABAC, Deception, and MTD. This study aims to increase the stress and cost for an insider to achieve unauthorized access. The key contributions of our paper are:

(1) introduce the concept of correlated attributes in ABAC.
(2) integrate deception into ABAC.
(3) extend ABAC model with moving target defense for insider threat mitigation and detection.

The remainder of this paper is structured as follows: Section 2 provides the background information. Section 3 summarizes the state of the art and related work. Section 4 describes the threat model. In section 5, we discuss the proposed framework. Section 6 describes the components of the proposed approach. In section 7, we discuss the experiment setup and results, while in section 8, we conclude and discuss the future work.

## 2 BACKGROUND

Access control systems (ACS) are critical components in the security architecture of organizations. Historically, traditional ACS, such as discretionary access control (DAC), mandatory access control (MAC), and role-based access control (RBAC) have successfully been used to protect organizations' resources. However, traditional ACS add more overhead and fewer security enhancements as the organization grows, leading to a need for advanced control models. ABAC is a higher-policy access framework that has drawn key attention in industry and academia. This work focuses on the National Institutes and Standards (NIST) ABAC framework [11] for consistency.

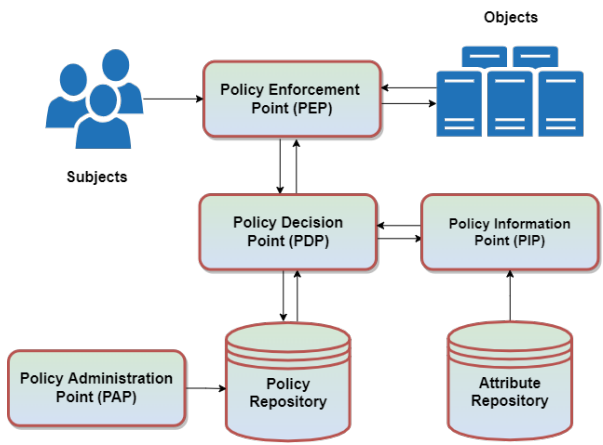

**Figure 1: The NIST ABAC Framework**

NIST describes ABAC as an access control model that authorizes subject's requests to carry out operations on objects based on the subject's attributes, object's attributes, environment attributes, and specified actions. The subjects can be either human or non-human (applications), such as automated software or legitimate bots. The attributes of subjects describe their properties and may include name, department, employee ID, branch, etc. In contrast, objects are mostly non-human, and their attributes may consist of creation date, type, department, and file size. Environment attributes describe the system's current state and may include geolocation, timestamp, weather condition, and threat level.

Intuitively, the NIST ABAC model shown in figure 1 has authorization policy rules which comprise subject attributes, object attributes, environment attributes, set of action(s), and the permission status, such as "grant" or "deny". The policy repository stores the policy rules, while the attribute repository stores the attributes. In addition, the model has four functional points connected in the pattern shown in figure 1. The functional points are: (1) Policy Administration Point (PAP); (2) Policy Information Point (PIP); (3) Policy Decision Point (PDP); and (4) Policy Enforcement Point (PEP). The PAP updates and manages the policy rules. The PIP manages access to the policy repository and attribute repository. The PDP analyses the attributes and policies to make logical access decisions for each access request. Lastly, the PEP enforces access decisions and ensures the subject does not go beyond the granted access decision.

## 3 STATE OF THE ART AND RELATED WORK

### 3.1 Moving Target Defense (MTD)

The US Department of Homeland Security describes MTD as "the concept of controlling change across multiple system dimensions in order to increase uncertainty and apparent complexity for attackers, reduce their window of opportunity and increase the costs of their probing and attack efforts" [19]. MTD involves proactively changing the configurations of a protected system to provide security guarantees. Ge et al [7]. assert that the configuration elements may change through shuffling, diversification, or distributed methods. In addition, the attack surface, detection surface, or both surfaces can be "moved" to incur more cost to the attackers. Oftentimes, security loopholes go unnoticed in critical infrastructures, which creates avenues for vulnerabilities that attackers can exploit. Although traditional approaches such as firewalls, intrusion detection systems, and vulnerability patching can be used to correct such security anomalies. However, because these approaches are reactive, attackers may gather knowledge about the system through reconnaissance and then launch attacks on the system before necessary corrections are made. MTD helps to resolve these vulnerabilities by limiting exposure to attack. MTD achieves this by: (1) negating any advantages that an attacker may have, in contrast to traditional approaches that assume security configurations remain immutable; (2) increasing the number of system's components attackers needed to compromise for successful attacks; and (3) increasing attack's time complexity. While achieving this goal, MTD ensures the usability of the access system are not affected by minimizing the response time and changes to end-user access patterns that the adoption of MTD may introduce.

### 3.2 Insider Threat

Insider threats have been studied for years and numerous strategies to reduce them have been put forth, including the use of deception, separation of duties and least privilege, behavioral analysis, and psychological analysis. For instance, Greitzer et al. [8] investigated the connection between employees' behavioral states and insider risks. These include disgruntlement, anger management issue, performance, stress, and aggressive behavior. Likewise, Hashem et al. [10] proposed detection of potential insiders using electrical signals generated by human biological activities, such as electroencephalography (EEG), electrocardiogram (ECG), and electromyography (EMG). Furthermore, Jiang et al. [15] proposed a model that combines deep learning and Graph Convolutional Networks (GCN) for detecting insider threats. Theoharidou et al. [26] presents an analysis of insider behaviour based on social science and criminology theories. Brown et al. [5] proposed an insider threat detection framework that investigates the relationship between the use of words and a set of risk factors that are either psychological or behavioral in nature. More recently, Paxton et al. [20] written and recorded incident notes to model insider attacks.

### 3.3 Deception Systems

Deception-based insider threat mitigation has garnered a lot of focus among cybersecurity professionals. Several works, both in industry and academia, have focused on using honey or decoys in various forms as the entity in systems to achieve the deception goal. Bercovitch et al. [2] suggested an "HoneyGen" system that mines rules to capture real data's features and generates fake data based on the rules. Similarly, Bowen et al. [4] introduced a Decoy Document Distributor (D3) model, which employs a rule-based approach to automatically generate and distribute decoy documents across a file system in order to entice malicious users. Bhagat et al. [3] proposed using honeypots to solve a network's intrusion detection problem. The authors investigated attacker interactions with honeypots and discovered that TCP [1] could be easily compromised among other network protocols. Srinivasa et al. [25] proposed using honey token fingerprinting as a decoy in a network system. Yuill et al. [27] proposed honeyfiles, an intrusion detection element that resides in a network server to bait hackers. Kaghazgaran et al. [17] introduced honey permissions and honey objects as extensions to role-based access control (RBAC) model for detecting insider threats. Similarly, Juels and Rivest considered various attack scenarios involving password theft. These include easily guessable passwords through brute-force, password compromise, and duplicity of passwords in systems. To mitigate such attacks, the authors proposed using honeywords to prevent attackers from guessing real passwords from a list of hashed passwords. Finally, Jaffarian et al.[14] proposed using a Multi-dimensional Host Identity Anonymization approach to defeat skilled attackers. The study suggest that reconnaissance attacks can be mitigated by modifying host addresses and fingerprints, anonymizing host fingerprints, and deploying honeypots and context-aware content.

## 4 PROBLEM STATEMENT AND THREAT MODEL

ABAC collects and analyzes attributes belonging to objects, subjects, and environment entities involved in a request to make access decisions. These attributes are assigned by dedicated attribute sources, which may be internal or external to an organization. We assume that these attribute sources are primarily outside of organizations' control, and therefore, there is the possibility of errors or vulnerabilities in the attributes. In this case, the incorrect assignment of attributes to access entities may adversely impact the reliability of the access control system. Based on this issue, we consider a threat model in which an insider can compromise the attributes in a given policy, for example, by creating an unauthorized attribute-access entity assignment or intentionally manipulating the attribute: value set or policy rules. An insider may compromise a given attribute through: (1) an unintended software error; (2) attribute forgery; or (3) creating and assigning new attributes to entities. In addition, actors in access control systems, such as subjects and objects, may exhibit new attributes. Here, we are extending our previous work by changing "moving" system configurations to mitigate the effect of attribute vulnerabilities and accommodate attributes' dynamic nature. In our previous work [1], we presented a defensive strategy for attribute-based access control that included a deception mechanism. We integrated a sensitivity estimator, which assesses object sensitivities and identifies sensitive objects that should be considered as potential deception targets; a honey attribute generator, which creates honey versions of sensitive objects for deception purposes; and a monitoring unit, which analyzes the PDP's access decision and monitors the use of honey attributes to detect the existence of insiders. To consolidate and extend the deception framework, we introduce an MTD-based concept in this work, in which the entities of the ABAC engine are changed to increase the cost of compromise for the ABAC elements. As a result, our approach aims to innovatively mitigate the insider threat and achieve dynamic access control.

---

[1] Transmission Control Protocol

## 5 THE PROPOSED FRAMEWORK

The primary objective of this work is to mitigate insider threat detection by introducing deception components and MTD to the standard ABAC system. Unlike previous work, which integrates MTD into ABAC, we consider new attributes of access entities and changes in the values of existing attributes. We employ a frequent pattern growth (FPGrowth) association algorithm, a more efficient association analysis algorithm than the Apriori. We extend the standard ABAC model with MTD-based components to dynamically mutate the policy rules using attributes that correlate to the original subject's attributes and object's attributes.

### 5.1 Integrating Honey Elements into ABAC

In our previous work [1], we introduced three additional modules to ABAC: Sensitivity Assessment (SA), Honey Attribute Generator (HAG), and Monitoring Unit (MU) in addition to the standard ABAC components:

**Sensitivity Assessment:** The SA module estimates the sensitivities of every object's attribute in the system. The sensitivity of an object's attribute depends on the type of action that users can perform on the object and the number of authorized users that can perform those actions. Based on this, we estimated the sensitivity of open, edit, and delete actions using a probability measure based on Shannon information content.[22]. Our assessment shows that writable objects have higher sensitivity than read-only objects because they require more confidentiality and integrity requirements.

**Honey Attribute Generation:** The Honey Attribute Generator module employs the Genetic Algorithm (GA) concept to generate honey attributes. GA is a population-based algorithm that generates a new population from initial ones by initializing population elements and then successfully performing crossover and mutation operations to generate the final population. Based on this, we generated non-distinguishable versions of sensitive attributes. We: (1) seeded the algorithm with initial version of real attributes; (2) evaluated the fitness score of each individual based on the semantic similarity between real and honey attributes using GLOVE embedding [2]; and (3) generated new population as the honey attributes.

**Monitoring Unit:** When a user requests access to an object, the PDP evaluates the request, and grants access if the user is authorized. On the other hand, the PDP denies access if the user does not have authorization while the user is presented with honey attributes as bait. Then, the MU monitors every attempt of the unauthorized user to ensure that the honey attributes are accessed.

## 5.2 Integrating Moving Target Defense (MTD) into ABAC

ABAC policy has several entities that can be used to expand the existing policy set dynamically. Intuitively, entities in an access request exhibit new attributes different from those in the original policy. Hence, in our approach, we use MTD to increase the difficulty and cost of successful insider attacks. Conceptually, the insider attack will be unsuccessful if the insiders do not know the mutated policies. To achieve this, we: (1) identify attributes as the changing components of ABAC; (2) identify the correlations among the attribute: value pairs by evaluating the access requests to identify new attributes which correlate to the attributes in the original authorization policy set; and (3) gracefully mutate the policy rules with correlated attributes. With this, the mutated policy set will prevent unauthorized access if the original policy becomes compromised.

## 6 THE EXTENDED ABAC WITH MTD AND DECEPTION

In this section we describe our proposed framework. We will: 1) informally discuss our approach by defining the standard ABAC components and extensions introduced in this work; 2) discuss the FPGrowth algorithm; 3) discuss the attribute correlation module; and 4) discuss the policy mutation engine to the standard ABAC that uses correlated attributes to mutate the policy rules.

### 6.1 Informal Description

We added the following entities and functions to achieve an MTD-enhanced deception framework.

**Deception Components**

(1) **Sensitive Attributes (**$SA$**):** a subgroup of object attributes that have high criteria for secrecy or integrity, making them vulnerable to insider assaults. The set of sensitive attributes $SA$ is defined as:
$SA = \{sa : sa \in OA \wedge Sensitivity(sa) \geq K\}$.

where $K$ is a predetermined threshold and $Sensitivity$ is a function used to calculate an attribute's sensitivity score. Each attribute $sa \in SA$ is a tuple: $sa = (name : value)$

(2) **Honey Attributes (**$HA$**):** a subgroup of attributes produced for every sensitive object as honey entities. Each attribute $ha$ is a member of the set $HA$ ($ha \in HA$) such that $ha$ is a tuple: $ha = (name : value)$

**MTD Components**

(1) **Support Threshold** ($SP_{\theta}$): the minimum value of support for association. We assume the heuristics for setting this value vary across organizations.

(2) **Confidence Threshold** ($CT_{\theta}$): the minimum value of confidence for association. We also assume the heuristics for setting this value vary across organizations.

(3) **Correlated Attributes** ($CA$): set of attributes generated through the correlation module. The set of correlated attributes $CA$ is defined as:
$CA = \{ca : ca \in CA \wedge Correlation(ca) \geq S\}$.
Where $Correlation$ is a function, that generates the attributes correlated to the original attributes and $S$ is a predefined support threshold. Each attribute $ca \in CA$ is a tuple: $ca = (name : value)$

(4) **Mutated Policies** ($MP$): set of mutated rules. Each mutated policy $mp$ is a member of the set $MP$ ($mp \in MP$).

**Functions**

To support our approach, we added the following extensions to the NIST ABAC model:

(1) **Correlation:** Returns a list of attributes correlated to the original attributes of particular object, as shown in the following:

$$Correlation : OA \rightarrow \mathbb{CA} \qquad (1)$$

Attributes which correlate to the original attributes have support and confidence values that are greater than the threshold values.

(2) **MutatedPolicy:** Generates a set of mutated policies using the correlated attributes. The function updates the original policy to the mutated policy set, as shown in the following:

$$MutatedPolicy : OP \rightarrow \mathbb{MP} \qquad (2)$$

The extended ABAC framework proposed is illustrated in figure 2. It has the main functional points of ABAC; PDP, PEP, PIP, and PAP. In addition, it has the sensitivity estimator and honey attribute generator as deception features. Further, we introduced the attribute correlation and policy mutation modules as MTD-based extensions. These extensions work together to: (1) identify correlated attributes; (2) mutate original policy using the correlated attributes; and (3) protect the system against insider attacks using the combination of honey data and mutated policies.

### 6.2 Frequent Pattern Growth

Frequent Pattern Growth, FPGrowth, is a technique for performing association analysis among elements in huge datasets. Association analysis is an approach for finding correlations or statistical relationships among data elements. Suppose a set of users have the same attributes of age, name, address, etc., but with unique IDs. Also, suppose these users frequently request access to objects which have the same values of attributes such as resource owner, date created, department, etc., but with unique IDs. Then, we can derive the frequent pattern for the subjects and object attributes in the data. Given that $X$ and $Y$ are sets of related subject attributes and object attributes, mathematically, the notation $X \implies Y$ denotes a statistical correlation between the two frequent sets. A natural question that follows this will be "How do we measure this association?". Support and Confidence are two metrics in the literature for measuring the strength of associations between $X$ and $Y$. Given an object dataset $D$ having $N$ total number of records. Suppose $X = \{X_1, X_2, ..., X_m\}$ and $Y = \{Y_1, Y_2, ..., Y_m\}$ are two non-empty object's itemsets in $D$, such that $X \neq \phi, Y \neq \phi$, and $X \cap Y \neq \phi$. If $P(X \cup Y) \neq \phi$ is the probability of $D$ records that contains both itemsets X and Y. Then, support of association between $X$ and $Y$, given in equation 3, is the percentage of records in $D$ that contains both itemsets $X$ and $Y$.

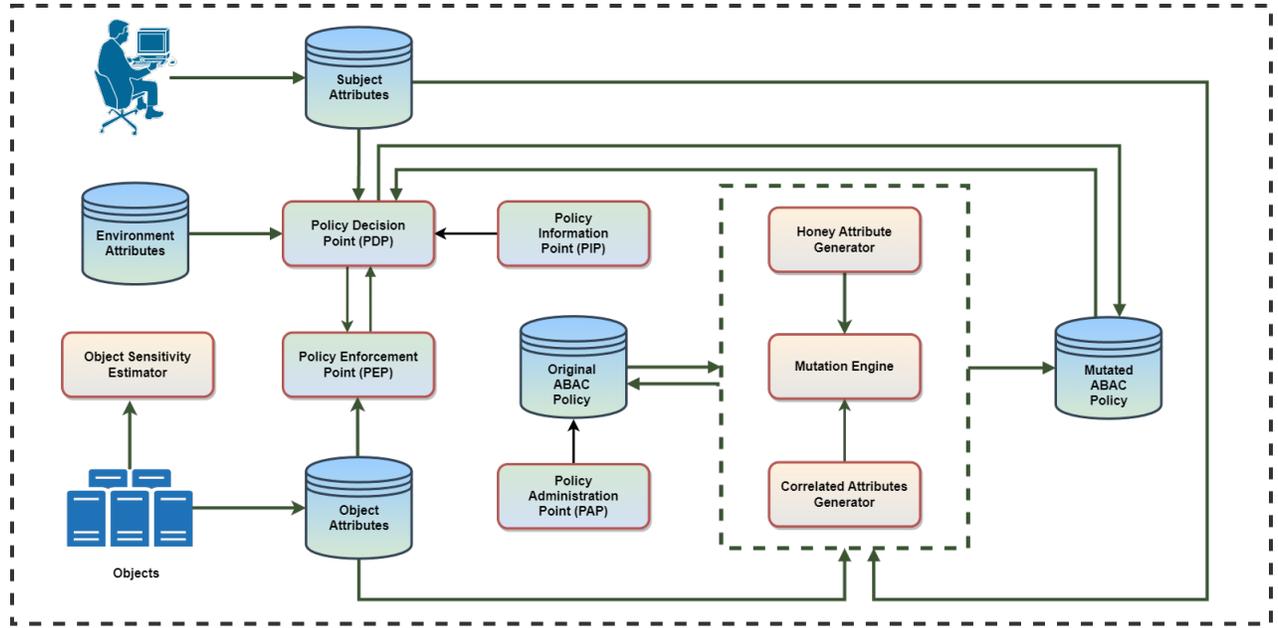

**Figure 2: The Attribute-based Insider Threat Mitigation Framework.**

$$Support(X \rightarrow Y) = P(X \cup Y) = \frac{Freq(X, Y)}{N} \quad (3)$$

Now, let us assume that $P(Y|X)$ is the conditional probability of having $D$ records that contains $X$ as well as $Y$. Then, confidence of association between $X$ and $Y$ given in equation 4 is the conditional probability between itemsets $X$ and $Y$. That is, the ratio of probability of $X$ and $Y$ occurring together to the probability of $X$ occurring alone. An association rule is said to be strong if the support and confidence values are greater than the respective threshold values.

$$Confidence(X \rightarrow Y) = P(Y|X) = \frac{Support(X \cup Y)}{Support(X)} \quad (4)$$

### 6.3 Attribute Correlation Module

We used the FPGrowth algorithm in subsection 6.2 to generate correlated attributes for the subjects and objects. We note that two common association algorithms, Apriori and FPGrowth can be used for this module. However, we chose the FPGrowth algorithm because it is more efficient than Apriori Algorithm which has high time complexity for two reasons. First, apriori generates candidate itemsets which grow as the dataset's size increases. Second, for each itemset generated, Apriori requires multiple scans of the dataset to verify the support. Therefore, apriori will be inefficient when memory is low and when the transactions are high. These limitations lead to increased time complexity.

On the other hand, FPGrowth uses a tree structure and a depth-first search pattern to register and mine all the frequent itemsets in the data. A frequent itemset is a set of items that usually appear together in a dataset. For example, if two or more objects have similar attribute-value pairs, then the set of attributes are frequent itemsets and can thus be regarded to have the same frequent structured pattern. Essentially, the FPGrowth algorithm scans the dataset two times only without generating a high number of candidate sets, thus reducing the search costs compared to the Apriori algorithm. Therefore, we gracefully used an FPGrowth algorithm to generate correlated attributes in our approach. We assume that environment attributes do not undergo frequent changes. Hence, we generated correlated attributes for the subject and object datasets only as described in the following steps: (1) We scanned the attribute records to identify all frequent items called (1-itemsets) and their frequencies (or support count). Then, we sorted the items based on support values in descending order and eliminated items with a support frequency less than the threshold; (2) We constructed the FPTree by scanning the dataset the second time. In this step, we created

---

**Algorithm 1:** Generation of Mutated Policy Set

---

**Input** : Access Request: $A_{Req}$,
Set of original policy: $P = \{r_1, r_2, ..., r_n\}$
**Output** : $MutatedPolicy = List()$

**1 Function** *getMutatedPolicy()*
// generates correlated attributes
**2**    $S_\theta$ = Support Threshold
**3**    $C_\theta$ = Minimum Confidence
**4**    $CorrAttributes = List()$
**5**    **for** $attr \in A_{REQ}.attributes()$ **do**
**6**      $Corr_{attr} = getCorrAttributes(attr)$
      // Get Support and Confidence
**7**      $Sup_{score}$ = getSupport($Corr_{attr}$)
**8**      $Conf_{score}$ = getConfidence($Corr_{attr}$)
**9**      **if** $Sup_{score} \geq S_\theta$ & $Conf_{score} \geq C_\theta$ **then**
**10**        $CorrAttributes.add(Corr_{attr})$
**11**      **end**
**12**    **end**
    // gets the list of attributes in the policy
     rules
**13**    $PolicyAttributes = List()$
**14**    **for** $r \in P$ **do**
**15**      $PolicyAttributes.add(getAttributes(r))$
**16**    **end**
    // Create Mutated Policies
**17**    $MutatedPolicy = List()$
**18**    **for** $item \in CorrAttributes$ **do**
**19**      $rule_{item} \leftarrow createRule(item)$
**20**      $MutatedPolicy \bigcup rule_{item}$
**21**    **end**
**22**    **return** $MutatedPolicy$

---

the Tree root. Then, we recursively generated a branch for each unique subject and object's record in the order of the List; (3) We mined the frequent pattern growth by identifying the lowest node, known as the suffix or frequent length-1 pattern. Then, in the FP-tree, we built its sub-data which is a set of prefix paths, also known as a conditional pattern, that coexists with the suffix pattern. Next, we constructed the conditional FP-tree by counting the number of itemsets in the sub-data with support greater than the threshold; and (4) We created the FPGrowth by combining the previous step's conditional FP-tree with the suffix pattern. The frequent pattern items in a smaller conditional FP-Tree constitute the correlated attributes.

### 6.4 Policy Mutation Module

We used the correlated attributes to generate a mutated policy set. In algorithm 1, we defined a *getMutatedPolicy* function for this purpose which involve three steps: (1) In lines 2 and 3, we defined the minimum support values for support and confidence. Then, we identified the attributes in each access request. This step is shown in lines 6; (2) For each attribute in the access request, we evaluated the correlated attributes and identified the ones that have support value and confidence value greater than the threshold values. This step is shown in lines $7 - 11$; (3) In lines 14 - 16, we collated the attributes in the original policy; and (4) We chose the correlated attributes to generate a mutated policy. This step is shown in lines $17 - 21$.

## 7 EXPERIMENT AND EVALUATION

### 7.1 Dataset

The effectiveness of an MTD approach depends on the number of components that change and the time of change. Since two actors, the subject and object, are in ABAC, we evaluated our approach on two separate datasets, one for the subject and another for the object while we assume that environment conditions are constant. We also assume that the training will achieve a good result if we obtain both subject and object datasets from the same sources. We obtained the subject and object attribute data from the California Basic Educational Data System (CBES), which was published by the Department of Education in 2018. The CBES collects data about students and staff across various schools and districts. We built the subjects' data in this study using tabular staff demographics and staff assignment datasets. Staff demographics data has 364,759 distinct records and 16 attributes, whereas staff assignment data contains 1,269,836 distinct records with 13 attributes. We integrated the two datasets using overlapping attributes to generate a single dataset with 1,269,836 distinct records and 23 attributes. Similarly, we obtained the object dataset from course enrollment data, which has 3,228,250 distinct object records and 23 attributes. In addition, we performed proportional sampling of 40,000 records from the generated datasets to lower the execution time. We can also evaluate the approach with synthetic datasets. However, the approach may not produce satisfactory results with synthetic datasets because the attributes may be skewed, resulting in low correlations.

**Table 1: Overhead of Attribute Correlation Module**

| S/N | Support Threshold | FPGrowth Time(s) | Apriori Time(s) |
|-----|-------------------|------------------|-----------------|
| 1 | 0.025 | 20.4622 | 154.4166 |
| 2 | 0.050 | 10.0144 | 132.7004 |
| 3 | 0.075 | 7.0509 | 127.8882 |
| 4 | 0.100 | 5.43514 | 127.2858 |
| 5 | 0.125 | 4.8038 | 127.2233 |
| 6 | 0.150 | 4.13404 | 126.9406 |
| 7 | 0.175 | 3.8763 | 126.5939 |
| 8 | 0.200 | 3.7909 | 125.4295 |
| 9 | 0.225 | 3.5830 | 125.1921 |
| 10 | 0.250 | 3.4153 | 124.7493 |

### 7.2 Implementation And Results

**Feature Selection:** We pruned the features with very low correlation using Dispersion Ratio measure, which finds the ratio of the arithmetic mean (AM) and the geometric mean (GM) of a feature. We removed the feature from the dataset if the ratio is lower than a specified threshold value.

**Time Overhead:** Table 1 shows the overhead added by the correlation module for both Apriori and FPGrowth algorithm. The

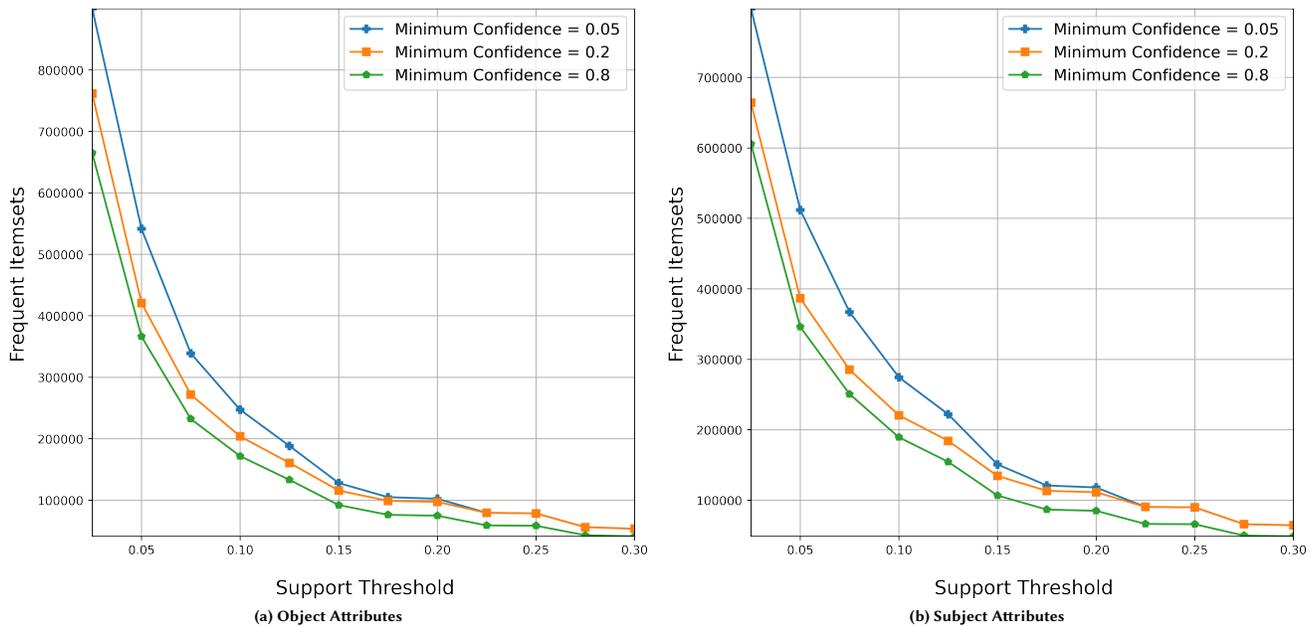

**Figure 3: Frequent Attributes Itemsets Against Support Threshold**

table reveals that that the time complexity reduces as the support threshold increases. This is because less attributes will be analysed as the support threshold increases. The table also reveals that FP-Growth has less overhead than Apriori algorithm. Therefore, we selected FPGrowth algorithm to generate the correlated attributes.

**Frequent Attributes Itemsets:** Figure 3(a) shows the frequent attribute itemsets generated by the FPGrowth algorithm for object dataset. Similarly, Figure 3(b) shows the frequent attribute itemsets generated by the FPGrowth algorithm for the subject dataset. The figure reveals that the frequent attributes itemsets reduce as the support threshold increases. This is because fewer attribute sets become frequent as the support threshold increases. Similarly, as the confidence threshold increases, the number of frequent attribute itemsets reduce.

## 8 CONCLUSION AND FUTURE WORK

In this paper, we proposed a framework for addressing insider threat challenges by incorporating moving target defense techniques and deception into the ABAC system. To accomplish this, we devised an algorithm that collated the original subject and object attributes in access requests, and generated correlated attributes for each subject and object attribute. Then, using the FPGrowth association algorithm, we mutated the original policy set with the correlated attributes to make it more difficult and costly for an insider to gain unauthorized access. The evaluation results revealed that the proposed approach effectively identified the correlated attributes and adequately generated a mutated policy set without affecting the access control systems' usability. In the future, we plan to explore additional methods such as deep learning-based approaches to generate the correlated attributes. We will also conduct a comprehensive user study to further evaluate the effectiveness and usability of the proposed approach.

## ACKNOWLEDGMENTS

This work was partially supported by the National Science Foundation under Grant No. 2006329.